\documentstyle[12pt,aasms4]{article}

\def\et{\em et al. \em}

\received{}
\accepted{}
\journalid{}{}
\articleid{}{}

\lefthead{}
\righthead{}

\begin{document}

\title{The local to global $H_0$ ratio and the SNe Ia results}

\author{Simon\,P.\,Goodwin, Peter\,A.\,Thomas, Andrew\,J.\,Barber, John\,Gribbin, Leslie\,I.\,Onuora}
\affil{Astronomy Centre, University of Sussex}
\affil{Falmer, Brighton, BN1 9QJ}

\begin{abstract}

The effects of differences between the local and global values of the 
Hubble parameter on the cosmologies consistent with studies of high-redshift 
Type Ia supernov\ae~are discussed.  It is found that with a local Hubble 
parameter around 10 per cent higher than the global value then open 
cosmological models (such as $\Omega_{M} = 0.3, \Omega_{\Lambda}=0$) 
are prefered and if the local value is around 20 per cent higher then 
standard cosmological 
models ($\Omega_{M} = 1, \Omega_{\Lambda}=0$) can be recovered.  Even in the 
case where the Hubble parameter ratio is 1, low $\Omega_M$ open 
cosmologies with $\Omega_{\Lambda} = 0$ are not rejected at the 95 per 
cent confidence level.

\end{abstract}


\section{Introduction}

Recent exciting results from the searches for high-redshift type Ia 
supernov\ae~(SNe Ia) have been interpreted as suggesting that there is 
indeed a cosmological constant and that the expansion of the Universe 
is accelerating (Perlmutter \et 1999 and Riess \et 1998).

The results on the cosmology are robust in that they are independent of 
the value of the Hubble parameter, $H_0$, (as they use redshift, $z$) and 
also of the local calibration of the SNE Ia luminosities (as the distance 
modulus is used).  They do rest upon an underlying assumption 
that the local and global values of $H_0$ are the same (so that 
the distance-redshift relationship does not change).    

Here we examine how different the local value of $H_0$ 
must be from the global value in order to significantly change the 
cosmology derived from the fitting of high-redshift SNe Ia 
data to the Hubble diagram.  A similar approach was taken by Kim \et 
(1997) to fit data from the first 7 high-redshift SNe Ia between 
$0.35 < z < 0.65$; in this paper 
we extend that work with a much larger number of supernov\ae~in the 
range $0.3 < z < 0.85$ .  (It should 
also be noted that the data used by Kim \et favoured an 
$\Omega_M = 1$ cosmology at the time.)

\section{Local and global differences in $H_0$}

The idea that the local Universe is atypical has been suggested by 
numerous authors (see Dekel 1994 for a review).

Hudson \et (1999) report evidence for a large scale bulk flow on scales 
possibly greater than 14000 km s$^{-1}$.  Plionis \& Kolokotronis (1998) 
show that there may be contributions to the X-ray cluster dipole from beyond 
16000 km s$^{-1}$.  Lauer \& Postman (1994) and Scaramella (1992) 
suggest the possibility of even larger density inhomogeneities out to a 
distance of 15000 - 30000 km s$^{-1}$.  

Phillips \& Turner (1998) see that near-IR galaxy counts out to 
$z=0.10 - 0.23$ are deficient which may be due to a local underdensity 
on such scales, implying that the local value of the Hubble 
parameter $H_0$ is up to 30 per cent higher than the global 
value.  Tammann (1998) also sees a decrease in the value of $H_0$ 
out to 18000 km s$^{-1}$ of 7 per cent.

Whilst these claims are not without their detractors it is clear that 
a body of work exists to suggest that the local value of $H_0$ may not be 
the same as the global value and could be higher. See Turner, Cen \& 
Ostriker (1992) for a detailed discussion of this topic.

\section{Method and results}

We assume that the value of $H_0$ on scales greater than $z=0.1$ 
is equal to the global value, and then fit open and flat cosmologies 
to the high-redshift SNe Ia data, shifting the zero-point of the data to 
find the best fit to the various cosmologies.  The difference in 
zero-points between the local and high-redshift supernov\ae~is then used 
to calculate the difference between the local and global values of the 
Hubble parameter.

Using the corrected data for 10 high-redshift SNe Ia presented in 
tables 5 and 6 of Riess \et (1998) and for 40 Supernova Cosmology Project 
(SCP) high-redshift SNe Ia from table 1 of Perlmutter \et (1999) we 
reconstruct the Hubble diagrams.  The shift in the local-to-global 
Hubble parameter required to fit various flat and open cosmologies 
are then calculated using a $\chi^2$ fit, also finding the 95 per cent 
confidence limits.

\placefigure{fits1}
\placefigure{fits2}

Figure 1 shows the required local-to-global ratio of the 
Hubble parameter to fit the SCP SNe Ia data.  The best fit when 
$H_{0(L)}/H_{0(G)} = 1$ in fig. 1(a) is the well-known 
$\Omega_M = 0.28, \Omega_{\Lambda} = 0.72$ cosmology found by 
Perlmutter \et (1999).

More interestingly a variety of low $\Omega_M$ open cosmologies are not 
rejected at the 95 per cent level even if $H_{0(L)}/H_{0(G)} = 1$.  In order 
for an $\Omega_M = 0.3$ open cosmology to be the best fit to the data 
then $H_{0(L)}/H_{0(G)} = 1.07$ is required.

The recovery of critical mass density cosmologies is more difficult. They are 
rejected at greater than the 99 per cent level if $H_{0(L)}/H_{0(G)} 
= 1$ and need a large ratio $H_{0(L)}/H_{0(G)} = 1.21$.

The data of Riess \et (1998) was also examined and the fits are very similar 
to those presented in fig. 1 although they are less 
significant due to the much lower number of data points.  Generally the 
Riess \et data are slightly less consistent with a low $\Omega_M$ 
open cosmology and $\Omega_M = 1$ critical cosmologies are ruled out at 
a higher confidence level, requiring a 1.2 to 1.3 $H_{0(L)}/H_{0(G)}$ 
ratio.  To fit best a low $\Omega_M$ open cosmology then the 
$H_{0(L)}/H_{0(G)}$ ratio must be 1.1 - 1.2.

\section{Conclusions}

If the local value of the Hubble parameter is higher than the global 
value on scales of a few 100 $h^{-1}$ Mpc then the fitting of cosmological 
parameters to the high-redshift SNe Ia data of Perlmutter \et (1999) 
and Riess \et (1998) may be inappropriate.  Previously popular cosmologies 
such as open or critical matter density Universes with no 
cosmological constant may be acceptable if we live in a local 
underdensity.   Whilst large underdensities of scales up to 300-600 $h^{-1}$ 
Mpc are not expected from standard power spectra, there is some 
observational evidence that we may live in such an underdensity (see 
section 2), a possibility which is not ruled out by these results.  It 
seems that is too early yet to abandon the traditional models.

\acknowledgements

LIO is a Daphne Jackson Fellow funded by the Royal Society.

\newpage

\figcaption[fits1.ps,fits2.ps]{The required ratio of local-to-global 
Hubble parameter to fit cosmologies of various $\Omega_M$ for SCP data in 
(a) a flat cosmology ($\Omega_M + \Omega_{\Lambda}=1$) and (b) an open 
cosmology ($\Omega_{\Lambda}=0$).  The 95 per cent confidence limits are 
marked in both cases by dashed lines.}

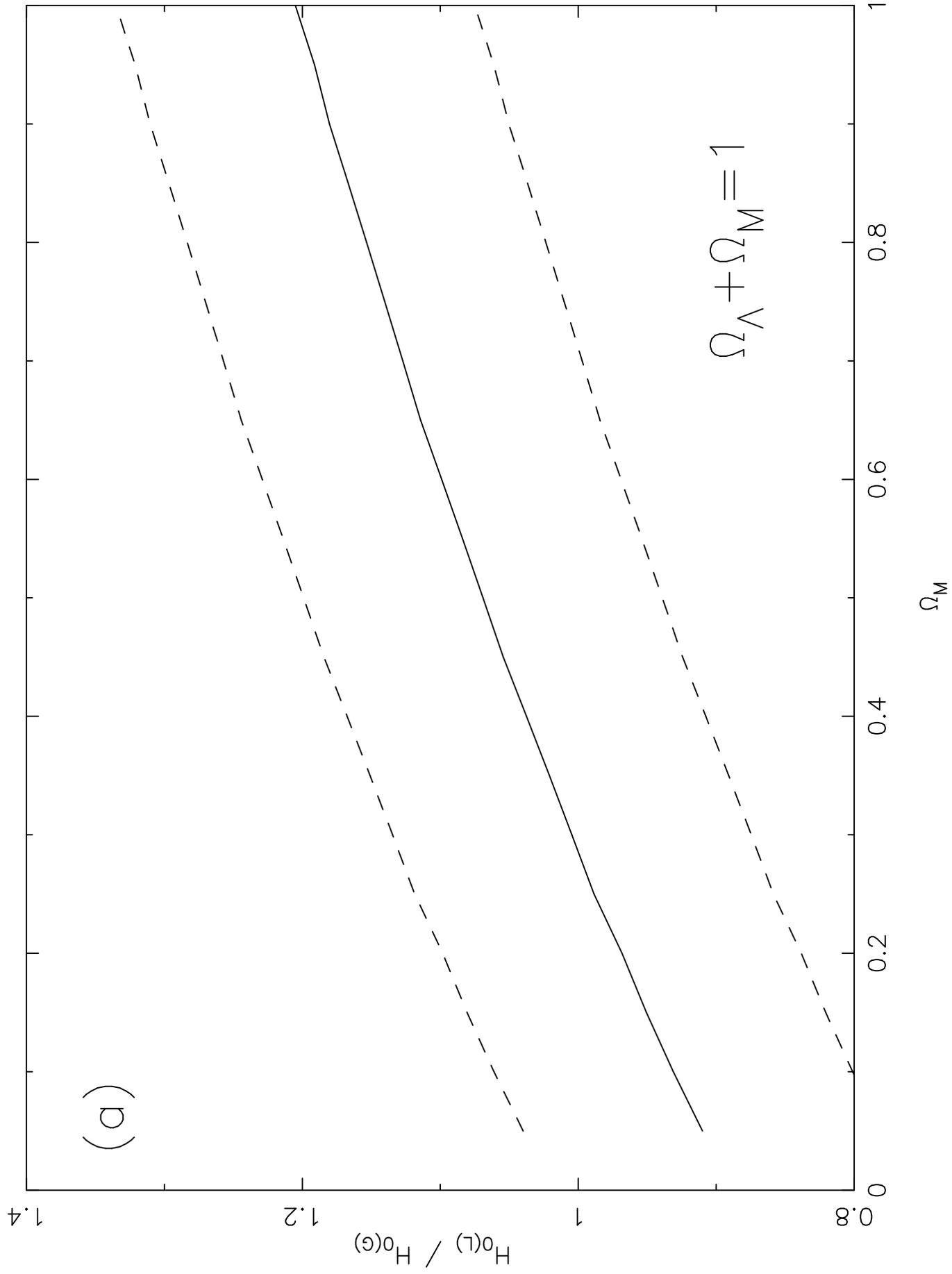
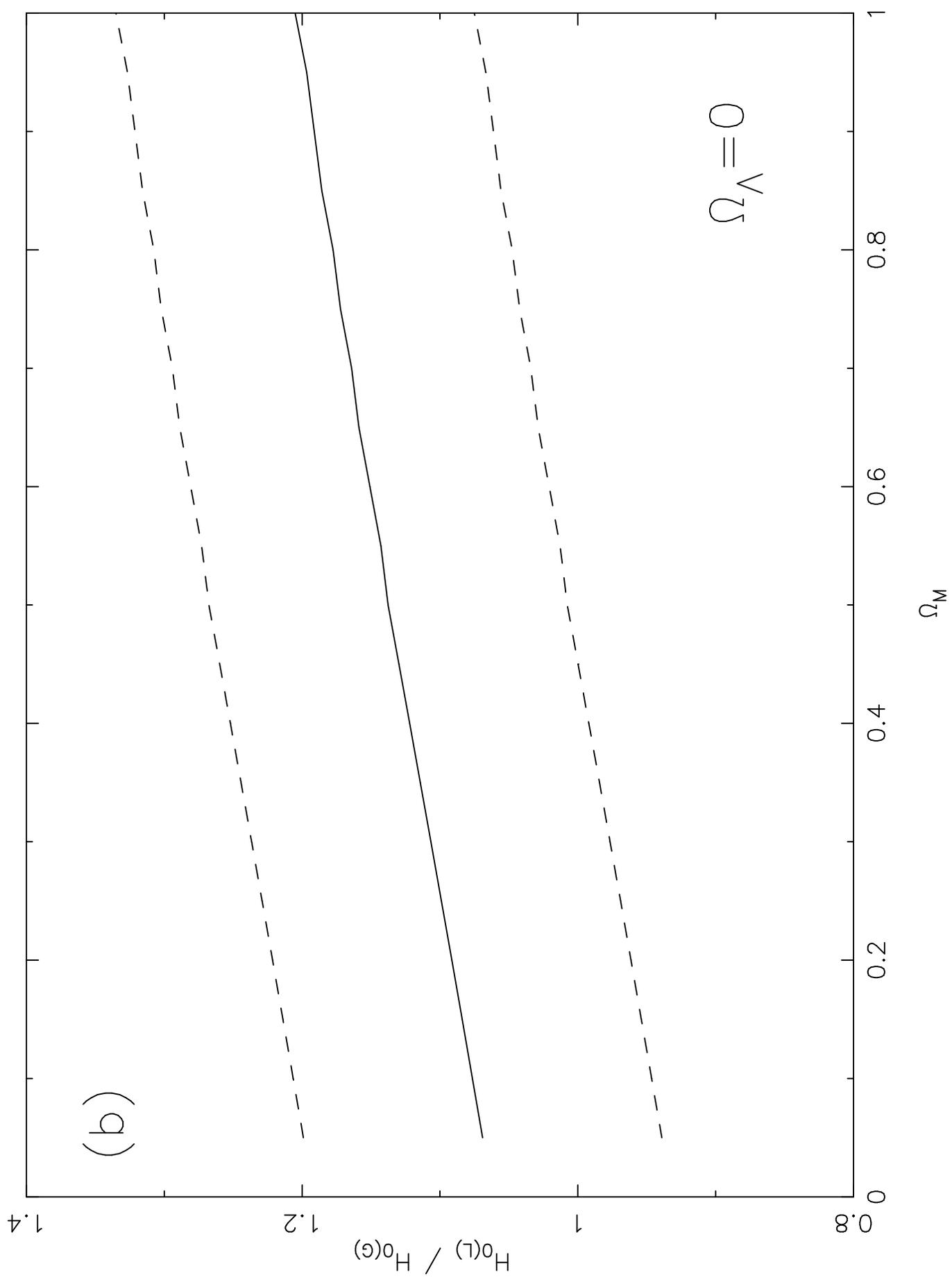
\end{document}